\documentclass[showpacs,preprintnumbers,amsmath,amssymb]{revtex4}

\usepackage{graphicx}

\def\be{\begin{equation}}
\def\ee{\end{equation}}
\def\bea{\begin{eqnarray}}
\def\eea{\end{eqnarray}}
\def\beq{\begin{eqnarray}}
\def\eeq{\end{eqnarray}}

\begin{document}

\title{A NOTE ON THE VIABILITY OF GAUSS-BONNET COSMOLOGY}
\author{R. Chingangbam}
\author{M. Sami}
\affiliation{Centre for Theoretical Physics, Jamia Millia Islamia,
New Delhi, India}
\author{P. V. Tretyakov}
\affiliation{Joint Institute for Nuclear Research, Dubna,
 Moscow Region, Russia.}
\author{A.V. Toporensky}

\affiliation{Sternbertg Astronomical Institute, Moscow State
University, University Prospect, 13, Moscow 119899, Russia}

\begin{abstract}
In this paper, we analyze the viability of a vacuum Gauss-Bonnet
cosmology by examining the dynamics of the homogeneous and
anisotropic background in 4+1 dimensions. The trajectories of the
system either originate from the standard singularity or from
non-standard type, the later is characterized by the divergence of
time derivative of the Hubble parameters for its finite value.
 At the onset, the system should relax to Einstein phase
at late times as the effect of Gauss-Bonnet term becomes negligible
in the low energy regime. However, we find that most of the
trajectories emerging from the standard big-bang singularity lead to
future re-collapse whereas the system beginning its evolution from
the non-standard singularity enters the Kasner regime at late times.
This leads to the conclusion that the measure of trajectories giving
rise to a smooth evolution from a standard singularity to the
Einstein phase is negligibly small for generic initial conditions.
\end{abstract}
\pacs{98.70.Vc}
 \maketitle

\section{Introduction}
Modified theories of gravity are under active consideration at
present in cosmology. Efforts are being made to mimic late time
acceleration from large scale modification of gravity without
resorting to exotic forms of matter dubbed {\it dark
energy}\cite{review,review1}. The extra dimensional effects can give
rise to modification of gravity; similar effects can be induced by
adding a generic function of Ricci scalar to Einstein-Hilbert action
giving rise to f(R) gravity (see Ref.\cite{fr} and references
therein). The quantum effects can also lead to higher order
curvature corrections to Einstein-Hilbert action. These corrections
can be systematically computed in perturbative regime of string
theory. Amongst all the higher derivative corrections which might
arise quantum mechanically, the Gauss-Bonnet (GB) correction has
distinguished features\cite{GB}. In this case, the equations of
motion continue to be of second order thereby ensuring the
uniqueness of their solutions. However, in 3+1 dimensions, the GB
term is topological in nature; it acquires dynamics only in higher
dimensions. Nevertheless, it can influence the 4 dimensional physics
if it is coupled to a dynamically evolving scalar field(s). The pure
GB term being in the higher dimensional bulk can also lead to
modification of Einstein equations on the brane\cite{Maeda}.

Attempts have recently been made to derive current acceleration
using the GB term coupled to a scalar
field\cite{NOS,KM06,CTS,Neupane,NO05,Cog06,Dede,Cal,Sami06,Annalen,Sanyal}.
The model exhibits remarkable property that it does not disturb the
scaling regime and can give rise to late time transition from matter
regime to late time acceleration\cite{KM06,TS}. This beautiful
result comes with a cost: the coupling of GB curvature invariant to
scalar field gets large at late times and can not be justified
within the perturbative regime the curvature corrections are
obtained; the model is also under pressure from nucleosynthesis
constraint\cite{KM06,TS}. On the theoretical ground, these models
are faced with other serious problems related to stability against
perturbations about FRW background\cite{Cal}. Similar situation is
expected to persist in the case of higher order Euler densities
coupled to scalar (dilaton/modulus) fields. Of course, one can argue
that these fields should be stabilized sufficiently early in order
to respect the nucleosynthesis constraints. It is, nevertheless,
important to examine the viability of Gauss-Bonnet cosmology in
general.

In this paper, we take a different route; we consider a vacuum 4+1
dimensional GB cosmology in a homogeneous and anisotropic background
and study the structure of generic singularities in the model.
Though 4+1 theories without compactification have no direct
applications to our Universe, study of their properties is important
for better understanding of gravity in four dimensions, showing its
specific properties in comparison with other cases. It is known, for
example, that in five dimensional Einstein gravity the uniqueness
theorem for a stationary black hole configurations is no longer
valid \cite{emparan}. Another classical example is related to the
disappearance of Mixmaster cosmological chaotic behavior in 10+1
dimensions \cite{Henneau}. These results have been formulated in the
framework of Einstein gravity. The Gauss-Bonnet term can further
modify traditional results known for 3+1 dimensional Einstein
theory. The main goal of the present paper is to study the
modifications of cosmological singularity due to Gauss-Bonnet term
in multidimensional cosmology.  In the low energy regime one might
expect the system to relax to 4+1 dimensional Kasner geometry.  We
shall examine the cosmological dynamics of the system under
consideration and investigate the measure of trajectories which
might connect to Einstein phase at late times.
\section{Evolution Equations}
We consider a $4+1$ dimensional theory with the action
\begin{equation}
S=\int \sqrt{-\mathrm{g}} (R + \alpha R^2_{GB})\,d^5x,
 \label{action}
\end{equation}

where $R^2_{GB}$ is the Gauss$-$Bonnet term

$$R^2_{GB}= R^{iklm}R_{iklm}-4R^{ik}R_{ik}+R^2.$$
In what follows we shall be interested  in the dynamics of the
system described by (\ref{action}) in the homogeneous and
anisotropic flat background with the metric
\begin{equation}
\mathrm{g}_{ik}=diag(-n^2(t),a^2(t),b^2(t),c^2(t),d^2(t)).
 \label{metric}
\end{equation}
This metric provides us a simplest modification of the standard
geometry allowing the realization of new dynamical regimes absent in
both Einstein gravity and  isotropic Gauss-Bonnet modified Einstein
theory of gravity (for a complete survey of possible 5-dimensional
cosmological backgrounds, see Ref. \cite{Hervik}).

It would be convenient to introduce Hubble parameters with respect
to four spatial dimensions $H_{a,b,c,d}=\frac{\dot a,\dot b, \dot
c,\dot d}{a,b,c,d}$. In the background described by the metric
(\ref{metric}), the action (\ref{action}) is a functional of of the
the scale factors and the lapse function along with their time
derivatives. Varying the action (\ref{action}) with respect to the
lapse function $n(t)$ and setting $n=1$ thereafter we find the
constraint equation
\begin{equation}
2H_aH_b+2H_aH_c+2H_aH_d+2H_bH_c+2H_bH_d+2H_cH_d+24\alpha
H_aH_bH_cH_d  =0,
 \label{3}
 \end{equation}
 which is the analogue of Friedmann equation in case of the geometry
 given by (\ref{metric}). Variation of (\ref{action}) with respect to scale factors leads to
the system of four dynamical equations,

\begin{eqnarray}
\begin{array}{l}
2(\dot H_b+H_b^2)+2(\dot H_c+H_c^2)  +2(\dot
H_d+H_d^2)+2H_bH_c+2H_bH_d+2H_cH_d + \\
\\+ 8\alpha\Big [ (\dot H_b+H_b^2)H_cH_d +(\dot H_c+H_c^2)H_bH_d
+(\dot H_d+H_d^2)H_bH_c\Big ] =0,
\end{array}
 \label{4}
\end{eqnarray}

\begin{equation}
\begin{array}{l}
2(\dot H_a+H_a^2)+2(\dot H_c+H_c^2)+2(\dot
H_d+H_d^2)+2H_aH_c+2H_aH_d+2H_cH_d +\\
\\+ 8\alpha\left [ (\dot H_a+H_a^2)H_cH_d +(\dot H_c+H_c^2)H_aH_d +(\dot
H_d+H_d^2)H_aH_c\right ] =0,
\end{array}
 \label{5}
\end{equation}

\begin{equation}
\begin{array}{l}
2(\dot H_a+H_a^2)+2(\dot H_b+H_b^2)+2(\dot
H_d+H_d^2)+2H_aH_b+2H_aH_d+2H_bH_d +\\
\\+ 8\alpha\left [ (\dot H_a+H_a^2)H_bH_d +(\dot H_b+H_b^2)H_aH_d +(\dot
H_d+H_d^2)H_aH_b\right ] =0,
\end{array}
 \label{6}
\end{equation}

\begin{equation}
\begin{array}{l}
2(\dot H_a+H_a^2)+2(\dot H_b+H_b^2)+2(\dot
H_c+H_c^2)+2H_aH_b+2H_aH_c+2H_bH_c +\\
\\+ 8\alpha\left [ (\dot H_a+H_a^2)H_bH_c +(\dot H_b+H_b^2)H_aH_c +(\dot
H_c+H_c^2)H_aH_b\right ] =0.
\end{array}
 \label{7}
\end{equation}
The evolution equations, in general, look cumbersome for analytical
investigations. In what follows we shall investigate the dynamical
regimes of the model numerically.

\section{dynamical regimes}

The presence of Gauss-Bonnet (GB) term allows some specific
dynamical regimes absent in pure Einstein gravity. First of all, the
volume of a flat Universe can have local extrema in this background.
The another new feature is associated with the possible existence of
a nonstandard singularity, found in Ref.\cite{Wheeler} (this type of
singularity was also found previously in another context in
Ref.\cite{Dem}, similar situation can also arise in 3+1-dimensional
cosmology with GB-term in presence of a dynamical dilaton
\cite{Kawai, Maeda, Tsujikawa}). Interestingly, the GB brane worlds
with the curvature term on the brane can also give rise to this type
of singularity\cite{Roy}. The  non-standard singularity, under
consideration, is characterized by $\dot H_i \to \infty$
($H_i=H_{a,b,c,d}$), for finite values of Hubble parameters. It
occurs when the major determinant of the system (4)$-$(7) vanishes.

We note that the generalized Kasner regime, being the solution of
vacuum equation motion for Bianchi I Einstein Universe, remains
intact in the low-energy regime, when Gauss-Bonnet contribution can
be neglected. This solution has the form $ds^2=-dt^2+\sum
t^{2p_i}dx_i^2$ with two known condition on the power indices
\begin{equation}
\begin{array}{l}
p_1^2+p_2^2+p_3^2+p_4^2=1,\\
p_1+p_2+p_3+p_4=1.
\end{array}
 \label{14}
\end{equation}

In the high-energy regime, the Gauss-Bonnet term becomes important.
However, in $4+1$ Universe there are no pure Gauss-Bonnet nontrivial
vacuum solutions, similar to found recently for the $5+1$
dimensional case \cite{TT}. To illustrate this point, le us consider
Eq. (3). We observe that there is only one term originating from the
Gauss-Bonnet contribution (the last term on the LHS), so this term
and the remaining Einstein contribution (first three terms of the
LHS) are equal in absolute values and opposite in sign. This means
that though a standard singularity (when all Hubble parameters tend
to infinity) is still possible, the dynamics in its vicinity is more
complicated than in $3+1$ Einstein or $5+1$ Gauss--Bonnet cases: the
Einstein and Gauss-Bonnet term are equally important in the $4+1$
case near a singularity. However, one particular regime can be
studied easily. If three Hubble parameters are equal and large
($H_a=H_b=H_c = H$), the equation (3) tells us that $H_d$ should
tend to zero near a standard singularity ($H_d=H/(1+4 \alpha H^2$),
and we have
$$
\dot H = - \frac{6H^2+36\alpha H^4+48\alpha^2
H^6}{3+48\alpha^2H^4+24\alpha H^2}, \,\,\,\,\,\,\,\,\,\, \dot H_d=-
\frac{-3H^2+12\alpha H^4}{3+48\alpha^2H^4+24\alpha H^2}.
$$

In this asymptotic regime the denominator is always positive, and we
can not meet the nonstandard singularity. Numerical studies confirm
that this singular regime can be smoothly matched with the
low-energy Kasner asymptotic.

%%%%%%%%%%%%%%%%%%%%%%
\begin{figure}
{ \centering

\includegraphics[width=10cm,height=8cm,angle=0]{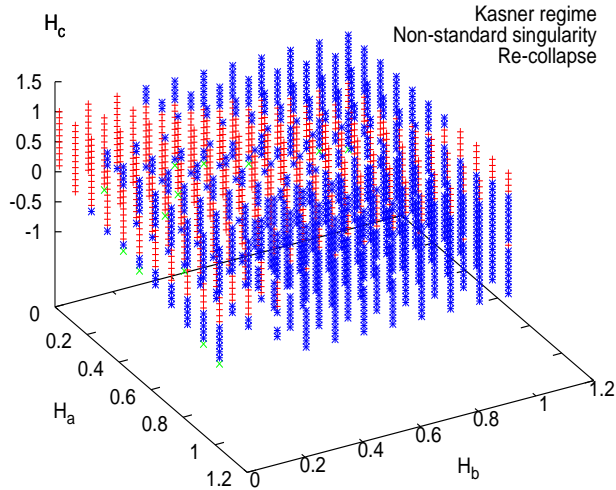}%,bb=0 0 504 720%     inpcomave.ps: 72dpi, width=17.78cm, height=25.40cm, bb=0 0 504 720}
\caption{The figure depicts a manifold of generic initial
conditions. With these conditions, the forward time evolution leads
to three possible outcomes, the low-energy Kasner regime,
re-collapse and nonstandard singularity. A negligible number of
trajectories correspond to nonstandard singularity.} }
\label{fig:1}

\end{figure}

\begin{figure}
{ \centering
\includegraphics[width=10cm,height=8cm,angle=0]{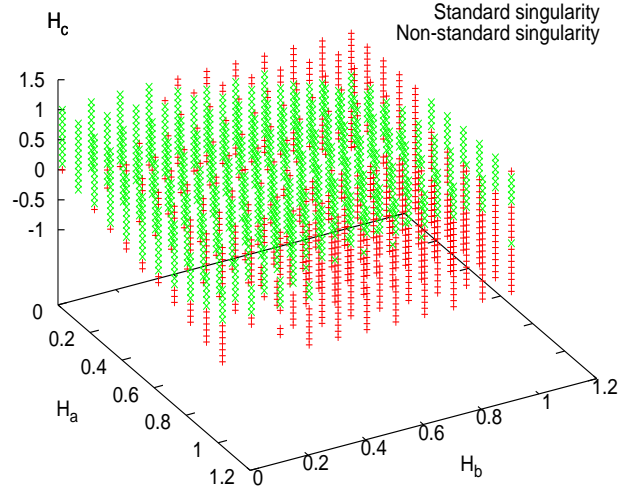}%,bb=0 0 504 720%     inpcomave.ps: 72dpi, width=17.78cm, height=25.40cm, bb=0 0 504 720}
\caption{Each point in the figure corresponds to a possible
trajectory. The backward evolution leads to two possible outcomes,
the standard singularity and the nonstandard singularity.
 The figure shows that around 40 percent of the trajectories originate from the standard singularity and the rest have their origin in the
  non-standard singularity.} }
\label{fig:2}

\end{figure}

\begin{figure}
{ \centering
\includegraphics[width=10cm,height=8cm,angle=0]{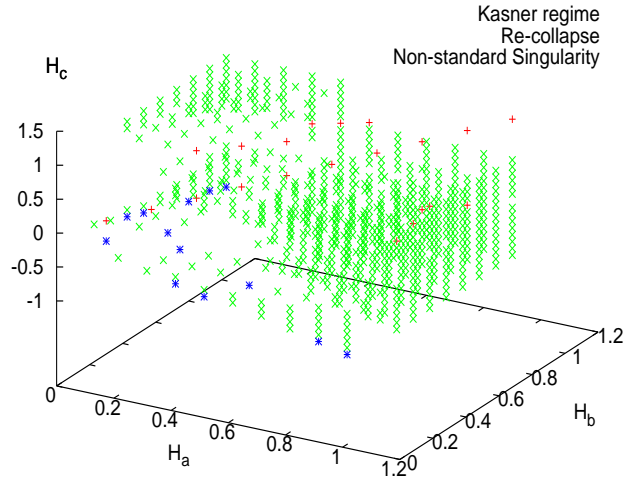}%,bb=0 0 504 720%     inpcomave.ps: 72dpi, width=17.78cm, height=25.40cm, bb=0 0 504 720}
\caption{Each point in the figure represents a trajectory originated
from the standard singularity. In forward evolution, different
trajectories evolve into low energy Kasner regime, non-standard
singularity and re-collapse. Around 95 percent of the trajectories
lead to re-collapse, about 1.5 percent encounter non-standard
singularity, and the rest fall into low energy Kasner regime.
Trajectories with three equal Hubble parameters go smoothly to
Kasner regime.} } \label{fig:3}
\end{figure}

\begin{figure}
{ \centering
\includegraphics[width=10cm,height=8cm,angle=0]{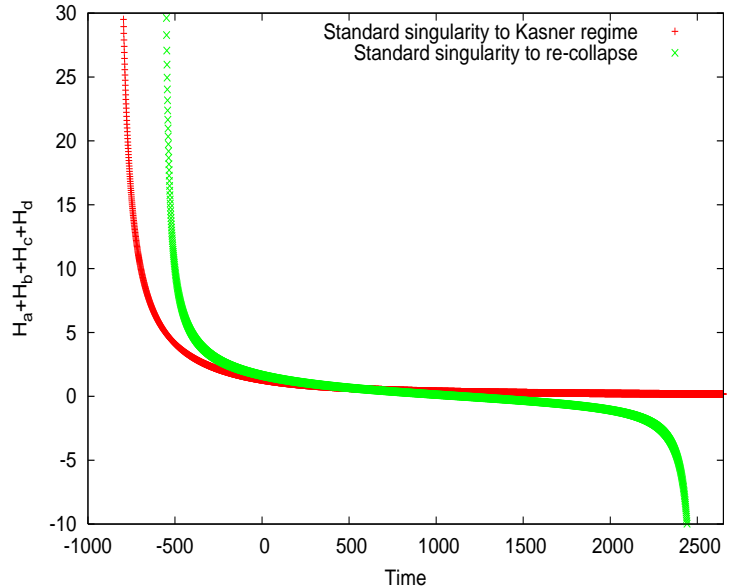}%,bb=0 0 504 720%     inpcomave.ps: 72dpi, width=17.78cm, height=25.40cm, bb=0 0 504 720}
\caption{The figure shows two typical trajectory evolving from
standard singularity to re-collapse and low energy Kasner regime,
the later being a rare possibility .} } \label{fig:4}

\end{figure}
%%%%%%%%%%%%%%%%%%%%%%
Interestingly, our numerical integrations show that it is the only
case when a Universe can evolve from a Big Bang singularity to
low-energy regime when Gauss-Bonnet contribution is negligible. All
other initial conditions lead either to trajectories originating in
the nonstandard singularity (or meeting this singularity in their
future evolution) or experience re-collapse back towards a
singularity.

Let us now spell out the details of the numerical  investigations of
the system described by the system of equations (1)-(7) for a large
set of initial conditions in the range $-1< H_i < 1$ (we fixed three
Hubble parameters and found the fourth from the constraint
equation). As we consider only positive values of the
coupling constant $\alpha$ in the present paper, we
set $\alpha=1$ in our numerical work. Our simulations reveal that: (a) During forward
evolution, around half  of the trajectories evolve to low energy
Kasner regime whereas in the case of the other half, we found that
the system re-collapses back to singularity. A negligible number of
trajectories lead to non-standard singularity in this case, see
Fig.1. (b) In the backward evolution, we distinguish two cases
representing two possible starting points of $4+1$-dimensional
universe, namely, a standard singularity and a nonstandard
singularity. We have found that more than 60 $\%$ initial conditions lead
to non-standard singularity; in 40 $\%$ cases, we find trajectories
evolving to standard singularity, see Fig.2.

We further observe interesting features while connecting the history
of universe  with its future evolution:
\begin{itemize}
\item For a particular set of initial condition with three equal Hubble parameters,
the system evolves to low curvature Einstein regime starting from a
standard singularity.

Except for the aforementioned particular case, we find that:
\item The trajectories originating from a standard singularity lead
to re-collapse for most of the initial conditions in their future
evolution(Fig.3 and Fig.4); the non-standard singularity is also
possible in rare cases.
\item Trajectories reaching a low-curvature Einstein regime  originate from a
nonstandard singularity.
\end{itemize}

As the measure of trajectories with three equal Hubble parameters is
zero, and our numerical results show that initial anisotropy of the
order of $10^{-5}$ is enough to destroy the smooth evolution and
that most of the trajectories reaching low-curvature regime must
have their origin in nonstandard singularity. Due to instabilities
of the numerical procedure near a singularity we can treat this
number as only an upper limit of possible anisotropy allowing a
smooth evolution. Similar result (no Big-Bang singularity) is
obtained  for the induced brane gravity with the Gauss-Bonnet
contribution in 5-dimensional bulk \cite{Roy}.

\section{Conclusions}
In the present paper we examined the dynamics of a flat anisotropic
4+1 dimensional universe in Gauss-Bonnet modified Einstein gravity.
The GB gravity in 4+1 dimensions in the FRW background leaves
standard big bang singularity unaltered. One could naively expect
that the dynamical system under consideration would settle to
Einstein phase in the low energy regime thereby leading to standard
description in 4 dimensional space time in the Kaluza-Klein
compactification scheme. It is really interesting that the
introduction of small anisotropy is capable of destroying the smooth
evolution and can lead to new dynamical regimes unknown to Einstein
or GB modified Einstein gravity in the homogeneous and isotropic
background. We have investigated the underlying dynamics of the
system numerically and observed several interesting features of the
dynamics. In case, $H_a=H_b=H_c$, our numerical results show that
trajectories starting from standard singularity can evolve to low
energy Kesner regime, however, the measure of such trajectories is
negligibly small. Excluding this particular case, we find that all
the trajectories beginning from standard singularity end up in the
non-standard singularity or encounter re-collapse in future. Within
this class of initial conditions, the trajectories which reach the
Einstein regime at late times are found to have their origin in the
non-standard singularity.

We thus conclude that the nonstandard singularity discussed earlier
in Ref.\cite{Wheeler} is not a very particular case of the dynamics,
it rather represents a typical feature of the cosmological dynamics
which frequently occurs during the evolution. The measure of the
system trajectories, which can smoothly connect the past standard
singularity with the low energy Einstein regime, is very small and
possibly can be zero. Further investigation are required to
understand whether this result is connected with specific properties
of 4+1 Gauss-Bonnet gravity or it is a more general feature of
Lovelock theory.

\section*{Acknowledgments}

The work of A.T and P.T was supported by RFBR grant 05-02-17450 and
scientific school grant 1157.2006.2 of the Russian Ministry of
Science and Technology. A.T. is grateful to the University of Jamia
Millia Islamia ,New Delhi, for hospitality where part of this work
was done. MS is supported by DST/JSPS (Grant No.
DST/INT/JSPS/Project-35/2007) and JSPS fellowship (FY2007). MS
thanks Gunma National College of Technology, Nagoya university and
Tokyo Institute of Technology where part of the work was done. RC thanks 
Tabish Qureshi for helping in numerical calculation.

\end{document}